\documentclass[%
 reprint,
 amsmath,amssymb,
 aps,
]{revtex4-2}

\usepackage{graphicx}
\usepackage{dcolumn}
\usepackage{bm}

\begin{document}

\preprint{APS/123-QED}

\title{Effects of the delocalized charge distribution in trapped ion-atom collisions}

\author{Ruiren Shi$^1$}
\author{Michael Drewsen$^2$}
\author{Jes\'us P\'erez-R\'ios$^1$}

\affiliation{$^1$Department of Physics and Astronomy, Stony Brook University, Stony Brook, New York, 11794-1901, United States
}

\affiliation{$^2$ Department of Physics and Astronomy, Aarhus University, DK-8000 Aarhus C, Denmark
}%

\date{\today}

\begin{abstract}
In the study of ion-atom interactions, the ion often remain trapped during the experiments. However, the effects of the trapping potential of the ion on ion-neutral interactions remain largely unexplored. Although trap-assisted ion-neutral complex formation has been experimentally studied~\cite{pinkas} and described by applying semiclassical theories where the ion is treated as a point charge particle~\cite{PhysRevLett.130.143003}, the potential effect of a delocalized charge distribution of a confined ion due to its quantum mechanical wavefunction has not been considered. To remedy this, in the present theoretical work we substitute the point charge of the ion with a delocalized charged distribution according to its motional ground state in the trap. Our results show that the trapping frequency and hence the spatial extension of the ion's ground-state wavefunction drastically affects the elastic and transport cross sections in interactions with neutral atoms. Stimulated by these results, we propose experimental procedures to verify the effects of the delocalize charge distribution in ion-atom interactions via measuring the heating rate of the ion due to the energy transfer in atomic collisions. Our novel approach brings new possibilities for investigating ion-neutral systems and, through them, new perspectives on ionic polarons and potentially a better understanding of trap-induced losses in ion-neutral experiments.

\end{abstract}

\maketitle


\section*{Introduction}

Charged-neutral interactions play an essential role in chemistry and biochemistry. In biology-relevant systems, charged-neutral interactions appear as part of complex reaction networks, mainly in solution, making it difficult to isolate and study their fundamentals~\cite{Bio1,Bio2}. Complementarily, combining ultracold atoms and cold ions opens up a new avenue to explore the fundamental properties of charged-neutral interactions~\cite{COTE20166,Hudson2019,RevModPhysatomion,LOUS202265,Review2024}. By controlling the internal state of molecules, it is possible to study state-dependent molecule-ion scattering processes~\cite{Michael2012,Michael2014,Henrik2022,Hudson2019,Stefan2012,Deiss2024}, and, with it, the nature of the underlying charged-neutral interactions. Similarly, the ion-atom interaction can be manipulated via Feshbach resonances when a single ion is placed in a sea of ultracold atoms~\cite{Weckesser2021}, opening possibilities to control atom-ion interactions~\cite{Schaetz2025}. The same scenario, under specific conditions, will give rise to the formation of ionic polarons--a quasi-particle appearing when the ion-atom interaction is dressed by the neutral bath, revealing even more intriguing properties of charged-neutral interactions~\cite{astrakharchik2020ionic,Bruun,Michael,Saajid2024}. 

An important means for such studies is a confining potential of the ion. Due to Earnshaw’s theorem, it is impossible to use static fields to trap a charged particle. Hence, either a combination of static electric and magnetic fields (Penning traps~\cite{Gosh1995}]), focused laser fields (optical dipole traps~\cite{Schneider2010}) or a combination of static and time-varying electrical fields (Paul traps~\cite{Gosh1995}) are applied. The latter method is most commonly used due to its simple construction and deep trapping potential, which for many purposes can be approximated by an effective harmonic potential in all directions. However, recently, it has been shown that the time-dependent trap induces the formation of transient atom-ion complexes. These complexes may act as a catalyst toward the formation of molecular ions via three-body recombination~\cite{Henrik2023}, compromising the stability of the ion. Similar effects governed by the effective harmonic confinement of the ion have been reported in the spin-flip transition between the ion and the atom~\cite{Ozeri2023}. In addition, it has been shown that ion-atom collisions in the presence of a trapping potential exhibit traces of chaotic scattering, revealing its key role~\cite{pinkas2024chaoticscatteringultracoldatomion}. Therefore, the trapping potential effectively modifies charge-neutral scattering properties, masking the fundamental free charge-neutral interaction. Similarly, the time-dependent trapping potential complicates the development of full quantum treatments in ion-atom systems since the trapping potential couples the relative and center of mass degrees of freedom, so the two-body problem behaves like a three-body problem. In addition, the strong long-range nature of the ion-atom interaction and the usually deeply bound molecular ion potentials require many channels to describe any meaningful scattering property. 

\begin{figure*}[t!]
    \centering
    \includegraphics[width=1\linewidth]{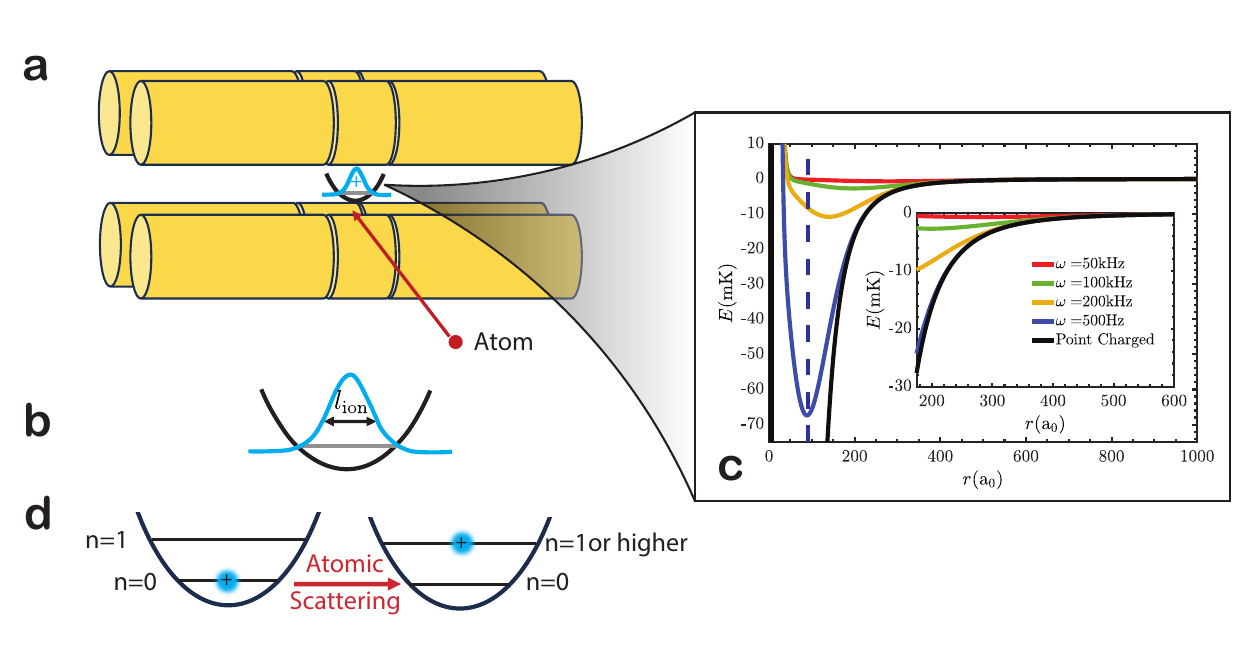}
    \caption{Sketch of an atom-ion scattering experiment. Prior to a scattering event, an ion is cooled to its quantum mechanical ground state in a 3D effective isotropic and harmonic trap potential (a). Due to the spatial extension of this ground state, the ion represents 3D isotropic Gaussian charge distribution, characterized by the total charge of the ion and the length scale $l_{\text{ion}}=\sqrt{\frac{\hbar}{m_{\text{ion}} \omega}}$, where $m_{\text{ion}}$ is the mass of the ion and $\omega$ the isotropic oscillation frequency (b). Panel (c) displays the ion-atom interaction potentials as a function of the trap frequency. The dashed line corresponds to the harmonic oscillator length scale, $l_{\text{ion}}$ for the $\omega=$500~kHz; see text for details. The inset shows a zoom-in of the long range region. At sufficient energy transfer in an atom-ion scattering event $E_{\text{trans}}\gtrsim \hbar \omega$, such an event will lead to motional excitation of the trapped ion to the state n=1 or higher (d), which can be detected by a motional state-dependent “digital” fluorescence signal through the application of the so-called shelving technique ~\cite{Diedrich1989,Deslauries2006}  (See main text for details).}
    \label{fig1}
\end{figure*}

This work proposes a novel viewpoint on trapped ion-atom collisions, as sketched in Fig.~\ref{fig1}. The ion can be brought to the ground state of the trapping potential. According to quantum mechanics, the ion will be delocalized following the ion's spatial wavefunction, consequently leading to a stationary finite spatial distribution of the ion's charge when in a motional eigenstate, like the ground state of the trapping potential considered here. The quantal elastic scattering, using two different atoms, shows a strong dependence on the trapping parameters, depending on the collision energy range being explored. Similarly, for realistic variation of experimental trap parameters, we find that in some range of collisional parameters, the ion diffusion and viscosity cross-sections, as well as the averaged energy transferred to the ion in a single collision, all present features that differ by about an order of magnitude or more with respect to the ion-trap frequency. Based on these findings, we discuss an experimental method that could measure the trapping potential's effect on ion-atom scattering by monitoring the ion heating rate.

\section*{Results and discussion}


\subsection{Trapped ion-atom interaction potential}
We consider a trapped ion of mass $m_{\text{ion}}$, in the ground vibrational state of the trap. For simplicity, we assume that the trap is isotropic and it is described by the trap frequency $\omega$. Hence, the ion is delocalized over a distance characterized by $l_{\text{ion}}=\sqrt{\frac{\hbar}{m_{\text{ion}} \omega}}$, as sketched in panel b of Fig.~\ref{fig1}, where $\hbar$ is the reduced Planck constant. The approaching atoms see the delocalization of the ion as a charged distribution, given by
\begin{equation}
\label{eq1}
\rho(R)=e|\Psi(R)|^2,
\end{equation}
where $e$ is the electron charge and $\Psi(R)=\left(\frac{1}{\sqrt{\pi} l_{\text{ion}}} \right)^{3/2}\exp{(-\frac{R^2}{2l_{\text{ion}}^2})}$ is the ground state wavefunction of the trapped ion. The electric field due to the charge distribution is   
\begin{equation}
\label{eq2}
    \mathbf{E}(r,l_{\text{ion}}) = \frac{e}{4\pi\epsilon_0}\left[\frac{\text{erf}\left(r/l_{\text{ion}}\right)}{r^2}-2\frac{\exp{({-\frac{r^2}{l_{\text{ion}}^2}})}}{\sqrt{\pi}rl_{\text{ion}}}\right]\mathbf{u}_r,
\end{equation}
where $\epsilon_0$ is the vacuum permittivity, erf($x$) is the error function of argument $x$, and $\mathbf{u}_r$ represents the unit vector along the radial direction. The electric field induces a dipole moment in the atom proportional to the atom polarizability $\alpha$, so the interaction potential between the trapped ion and the atom reads as 
\begin{equation}
U(r,l_{\text{ion}}) = -\frac{e^2}{2(4\pi\epsilon_0)^2}\alpha E^2(r,l_{\text{ion}}).
\end{equation}
The trapped ion-atom interaction potential for different trapping frequencies is shown in panel c of Fig.~\ref{fig1}, as a function of the atom distance $r$ for $^{137}$Ba$^+$-$^6$Li. At long distances, $r\gg l_{\text{ion}}$, $E(r,l_{\text{ion}})=\frac{e}{4\pi\epsilon_0 r^2}$, thus, recovering the point charged electric field, and hence, the charged distribution-atom interaction potential behaves as the charged-induced dipole interaction $\propto r^{-4}$. Instead, for  $r\approxeq l_{\text{ion}}$, the charged distribution-atom interaction shows a long-range minimum, absent in the case of point charged-atom interaction. Finally, the total charged distribution-atom interaction potential is given by 
\begin{equation}
\label{eq4}
V(r,l_{\text{ion}})=\frac{C_8}{r^8}+U(r,l_{\text{ion}})
\end{equation} 
where $C_8$ represents the short-range interaction coefficient due to the overlap of the electronic clouds of the ion and the atom, characteristic of atomic interactions.

Since the above model is based on the ionic wavefunction not being significantly disturbed during the collision, it is already here worth mentioning that for a collisional energy corresponding to a temperature of $\sim$1 mK for a light atom as He or Li, we estimate the collisional time to be of $\sim$10 ns, which is significantly shorter than the classical oscillation period of the trapped ion, and hence we expect not the ion's wavefunction to be perturbed significantly during the collision. However, for the lowest collisional energies considered in this paper, this may not turn out to be the case, and corrections to the model will have to be applied.    

\subsection{Atom-ion elastic cross section}
\begin{figure}[h!]
    \includegraphics[width=1\linewidth]{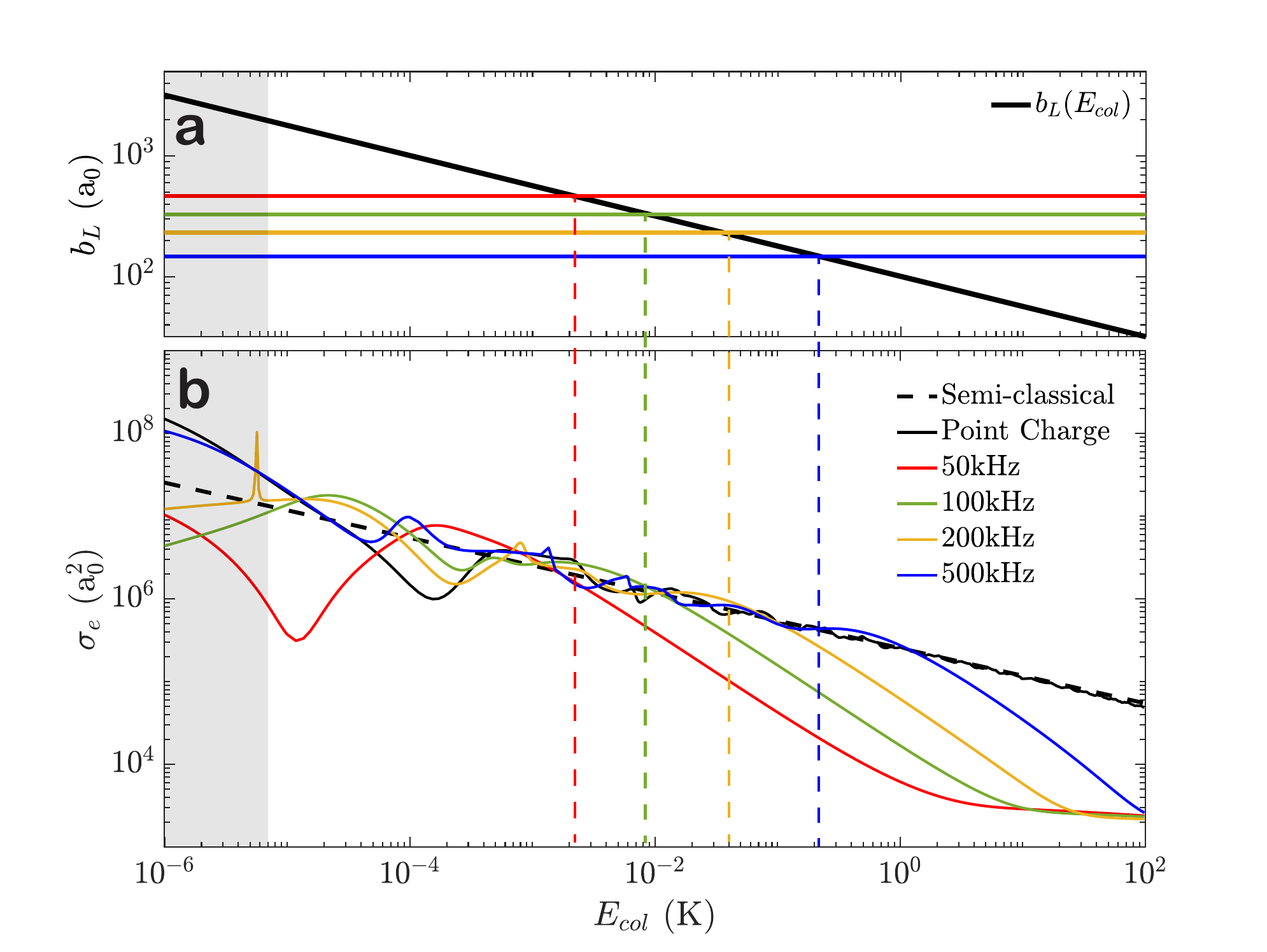}
    \caption{Elastic cross section for $^{137}$Ba$^+$-$^6$Li collisions in the presence of a trap. Panel a  displays the relevant length scales for the system under consideration: the Langevin impact parameter, $b_L$, and $l_{\text{distrib}}$. The solid horizontal lines represent $l_{\text{distrib}}$ for different frequencies (color code, see panel b). Panel b shows the elastic collision cross section in Bohr radii squared as a function of the collision energy in Kelvin. The vertical dashed lines denote the regions where the charged distribution deviates from the point charge results. The shade gray area in both panels represents the s-wave scattering dominated area, or ultracold regime.  }
    \label{fig2}
\end{figure}

 We have calculated the elastic cross section for trapped ion-atom collisions as a function of the collision energy $E_{\text{col}}$. The results for $^{137}$Ba$^+$-$^6$Li collisions against the free ion-atom case are shown in Fig.~\ref{fig2}. The trapped ion results show very similar behavior as the point charge $\propto E_{col}^{-1/3}$ (see Methods). Depending on the trapping frequency, there is a particular collision energy for which the elastic cross section deviates significantly from the free ion case, illustrated by the dashed lines. Specifically, we found that loose traps, i.e., wide spatial charge distribution, deviate at lower collision energies than tighter traps (more localized charge distributions). Surprisingly enough, at higher collision energies (1-10$^{2}$~K), we notice that, independently of the trapping frequency considered, the elastic cross section is almost the same. At higher collision energies, the short-range region of the interaction potential dominates the scattering observables, and since the short-range is primarily independent of the trapping frequency, as it can be seen in panel (c) of Fig.~\ref{fig1} and from Eq.~(\ref{eq4}), yields almost the same elastic cross section.

\begin{figure}[h!]
    \includegraphics[width=1\linewidth]{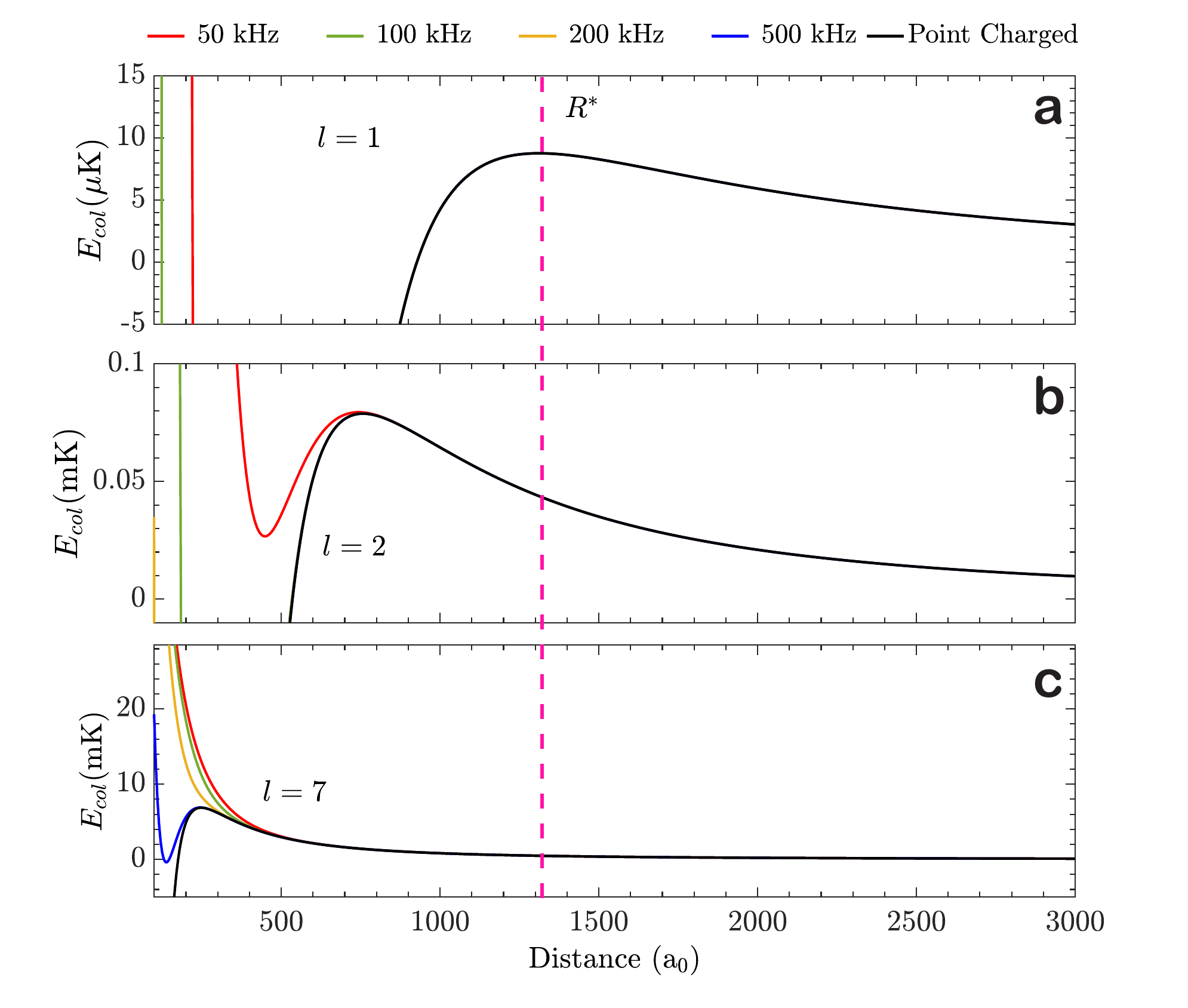}
    \caption{Effective single channel potentials for $^{137}$Ba$^+$-$^6$Li collisions. Panel a is for $l$=1 or p-wave, panel b is for $l$=2 or d-wave, and panel c is for $l$=7. The vertical dashed line represents the polarization length $R^*$ that characterizes the onset of ultracold scattering. }
    \label{fig3}
\end{figure}

In the case of a free ion, the range of the interaction potential relevant for a given collision energy is the Langevin impact parameter $b_L(E_{\text{col}})=\sqrt{\frac{2\alpha}{E_{\text{col}}}}$, which, from a quantum mechanical standpoint, specifies the largest partial wave contributing to the scattering. On the other hand, the trapped ion-atom interaction present depends on the typical length scale of a particle in harmonic oscillator $l_{ion}$, giving rise to an inherent length scale $l_{distrib}=l_{\text{ion}} \left( \frac{2}{\sqrt{\pi}} + \sqrt{\frac{3}{2}-\frac{4}{\pi}} \right)$ (see Methods). Therefore, when the Langevin impact parameter is similar to the inherent length scale of the trapped ion-atom interaction, i.e., $l_{distrib}\approx b_L(E^*)$, the atom-ion interaction potential becomes sensitive to the charged distribution of the trapped ion, yielding a different cross section from the point charged-induced dipole moment case. In other words, when $E_{\text{col}}\gtrsim E^*$, the elastic cross section shows a deviation from the free ion case as a consequence of the trapping potential. To support this idea further, we have calculated the effective potential, $V_l(r,l_{\text{ion}})=V(r,l_{\text{ion}})+\frac{l(l+1)\hbar^2}{2\mu r^2}$, defined for every partial wave $l$, and depending on the reduced mass of the ion-atom system $\mu$. The results are shown in Fig.~\ref{fig3}. First, all the trapping effects occur at $r<R*$, where $R^*=\sqrt{\frac{e^2 \alpha \mu}{\hbar^2(4\pi \epsilon_0)^2}}$ is the polarization length.  Therefore, the onset of ultracold scattering remains at the same collision energy, even in the presence of a trapping potential. For low partial waves, the trapping potential does not modify the long-range barrier, explaining why the trapped ion-atom elastic cross section at low collision energies is similar to the one for the free ion-atom case, as shown in Fig.~\ref{fig2}. On the contrary, at larger partial waves, the trapping potential drastically affects the position and height of the barrier, explaining the deviations observed in Fig.~\ref{fig2} for the elastic cross section between the trapped ion versus the free ion interaction with an atom.

\begin{figure}[h!]
    \includegraphics[width=1\linewidth]{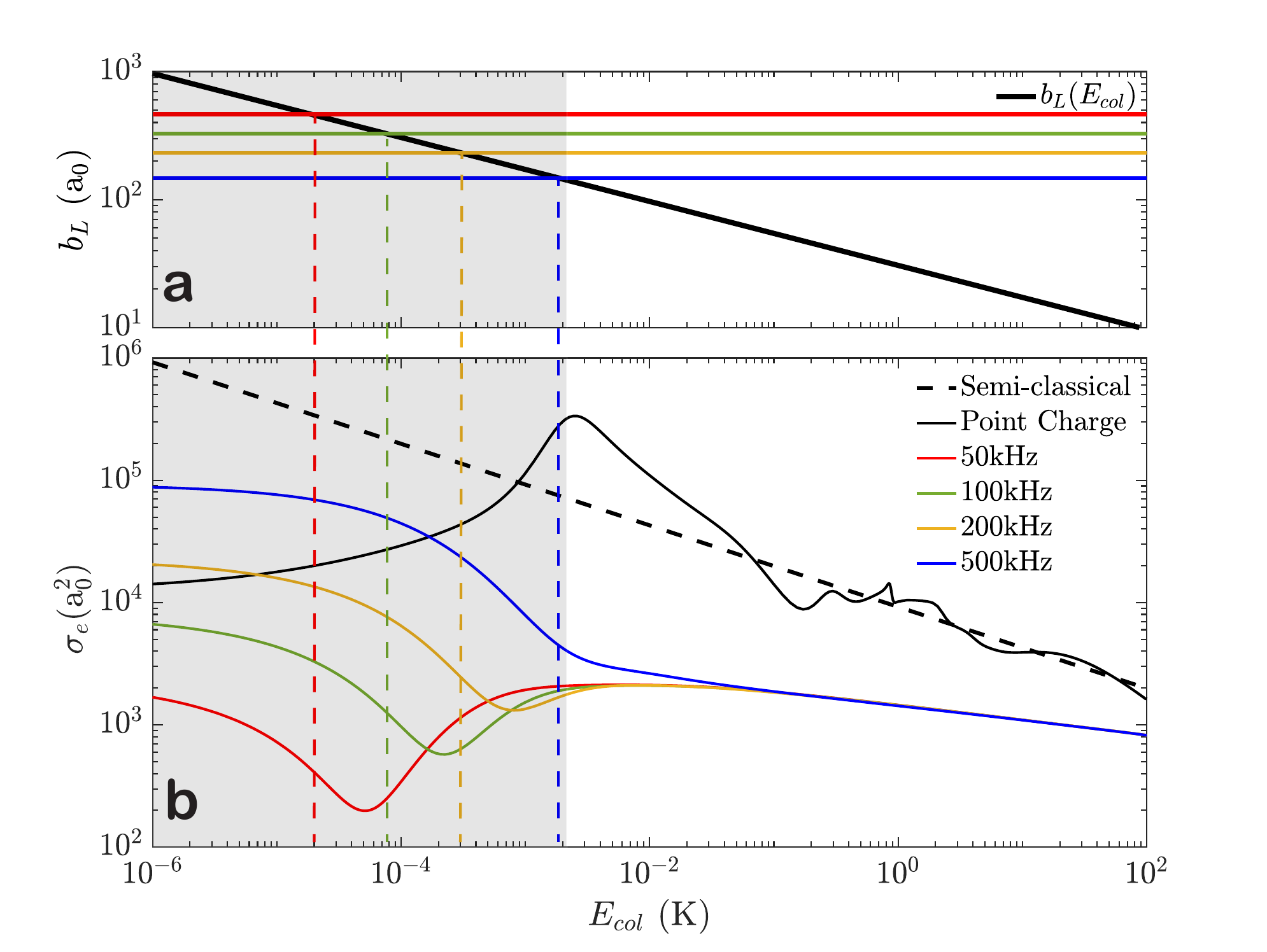}
    \caption{Elastic cross section for $^{137}$Ba$^+$-$^4$He collisions in the presence of a trap. Panel a  displays the relevant length scales for the system under consideration: the Langevin impact parameter, $b_L$, and $l_{\text{distrib}}$. The solid horizontal lines represent $l_{\text{distrib}}$ for different frequencies (color code, see panel b). Panel b shows the elastic collision cross section in Bohr radii squared as a function of the collision energy in Kelvin. The vertical dashed lines denote the regions where the charged distribution deviates from the point charge results. The shade gray area in both panels represents the s-wave scattering dominated area, or ultracold regime. }
    \label{fig4}
\end{figure}

The results for $^{137}$Ba$^+$-$^4$He collisions compared to the free ion-atom collision are shown in Fig.~\ref{fig4}. These are very different from the ones for $^{137}$Ba$^+$-$^6$Li displayed in Fig.~\ref{fig2}. In this figure, we observed a drastic effect of the trapping potential at all collision energies since $l_{distrib}\gtrsim R^*=98$~a$_0$ and the  
$l_{distrib}\approx b_L$ are clearly in the ultracold regime or s-wave dominated region. Therefore, the properties of the atom establish the role of the trapping potential on the onset of ultracold collisions. At high collision energies, as in the case of $^{137}$Ba$^+$-$^6$Li, $^{137}$Ba$^+$-$^4$He collisions show the same elastic cross section, characteristic of the same short-range interaction potential independently of the trapping potential. 

\subsection{Ion transport properties}

The trapping potential of the ion influences the ion-atom elastic scattering. Therefore, it may affect the transport properties of the ion, such as diffusion, viscosity, and, ultimately, its mobility in a neutral media. The diffusion properties of ions in neutral environments are characterized by the diffusion cross section, $\sigma_d$. In contrast, the viscosity properties are encapsulated in the so-called viscosity cross section $\sigma_{\eta}$ (see Methods). Both of these so-called transport cross sections have been calculated as a function of the trapping frequency using two different atomic baths.

\begin{figure}[h!]
    \includegraphics[width=1\linewidth]{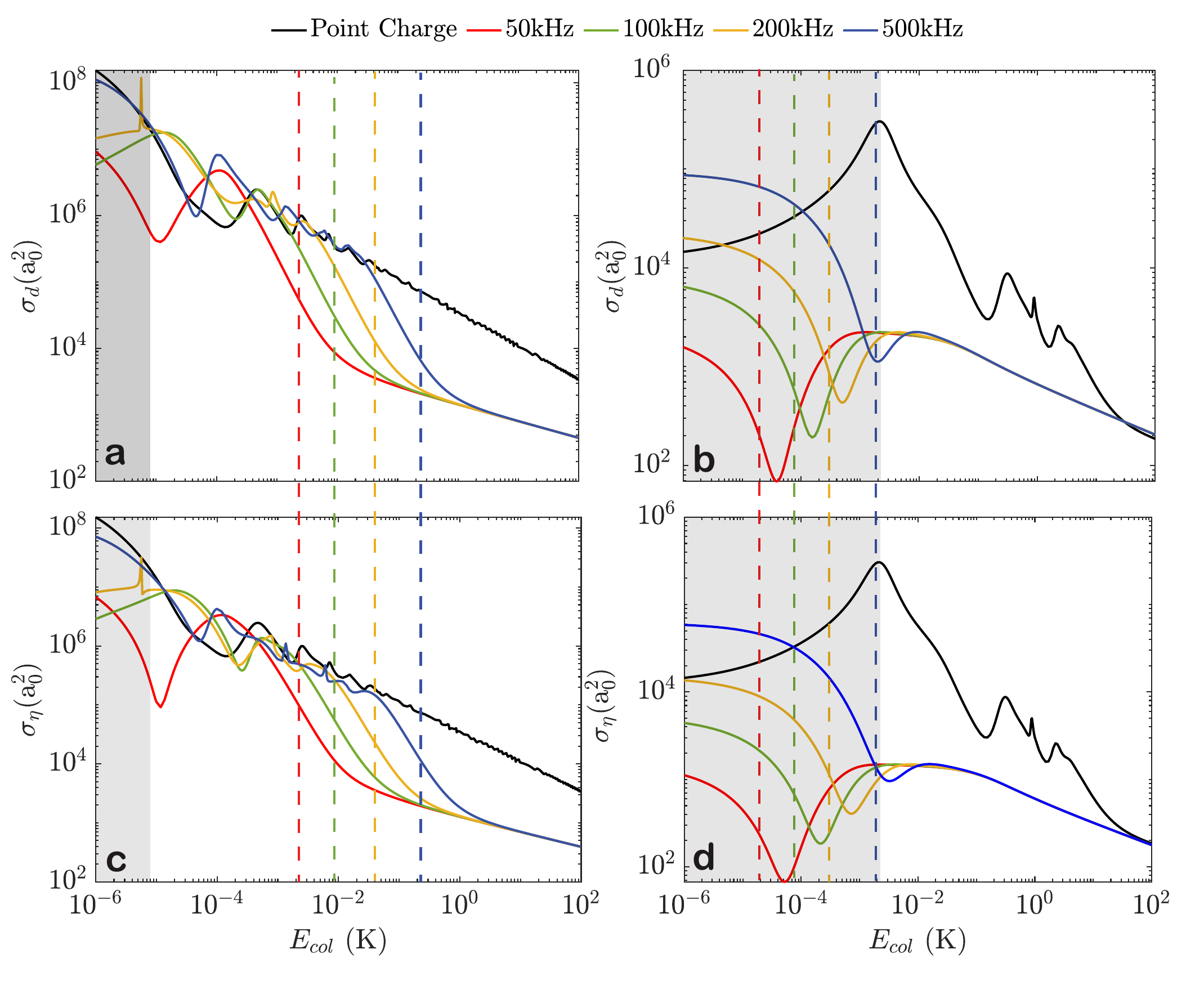}
    \caption{Atom-ion diffusion and viscosity cross sections as a function of the collision energy. Panels a and b display the diffusion cross section for $^{137}$Ba$^+$-$^6$Li and $^{137}$Ba$^+$-$^4$He collisions, respectively. Panels c and d depict the viscosity cross section for $^{137}$Ba$^+$-$^6$Li and $^{137}$Ba$^+$-$^4$He collisions, respectively. The color code in every panel is the same according with the legends in the upper part of the figure. }
    \label{fig5}
\end{figure}

The results for the diffusion and viscosity cross sections as a function of the collision energy for $^{137}$Ba$^+$-$^6$Li and $^{137}$Ba$^+$-$^4$He collisions are displayed in Fig.~\ref{fig5}. Overall, the behavior is very similar to the elastic cross section (Figs.~\ref{fig2} and \ref{fig4}), as long as $E_\text{col}\lesssim E^*$ the transport cross sections agree with the free ion-atom trend. On the contrary, $E_\text{col}\gtrsim E^*$ the transport cross sections show a more steep change than the elastic cross section, experiencing a change of more than one order of magnitude in only one decade of collision energy. For $^{137}$Ba$^+$-$^4$He, due to the very small polarization length, the trap readily affects the scattering properties of the system at any collision energy, as for the elastic cross section (see Fig.~\ref{fig3}). However, the diffusion and viscosity cross sections at high collision energies are identical, agreeing with the free ion-atom interaction potential. At high collision energies, the short-range region of the interaction potential is the most influential for scattering observables. Therefore, the short-range for the ion-atom interaction is unaffected by the trapping frequency.

\begin{figure}[h!]
    \includegraphics[width=1\linewidth]{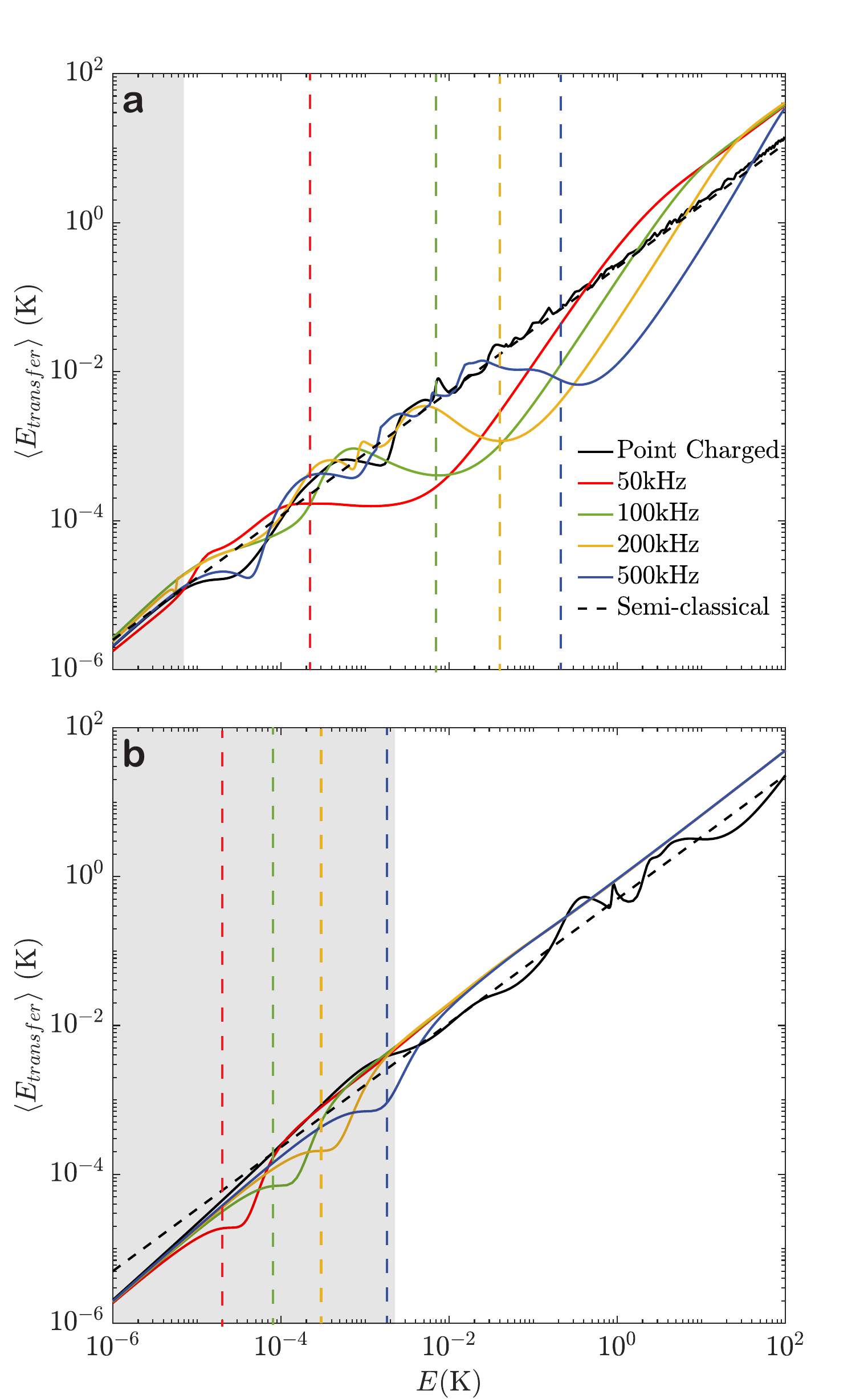}
    \caption{Averaged energy transfer per collision for trapped ion-atom collision as a function of the collision energy. Panel (a) display the results for $^{137}$Ba$^+$-$^{6}$Li collisions and panel (b) is for  $^{137}$Ba$^+$-$^{4}$He collision. The vertical dashed lines represent the borderline between regions where the charged distribution deviates from the point charge results. }
    \label{fig6}
\end{figure}

\subsection{Ion heating}

The results for the average energy transfer per collision energy for $^{137}$Ba$^+$-$^6$Li and $^{137}$Ba$^+$-$^4$He collisions for different trapping frequencies are shown in Fig.~\ref{fig6}. First of all, we notice, as expected from a semi-classical theory, that the energy transfer generally grows as $E_{col}^{5/6}$ (dashed line on the figure). However, there exist ranges of collision energies where $ E_{\text{transfer}}$ varies strongly with the trap frequency, when $E_{col}\gtrsim E^*$, indicating that at these collisional energies there is a large effect due to the spatial extension of the ion ground state wavefunction. For instance, for $^{137}$Ba$^+$-$^6$Li collisions this energy range is between 10$^{-3}$-1~K, the energy transfer various with up to $\sim$50 times with the trap frequency,  whereas for $^{137}$Ba$^+$-$^4$He collisions the range is 10$^{-5}$-10$^{-3}$~K, the variation is more like $\sim$5 times. Physically, the structures in $E_{\text{transfer}}$ originate from changes in the energy of shape-resonances due to the significant change in the ion-atom interaction potential with respect to the extension of the ion wavefunction (See Fig.~\ref{fig1}). While the exact energy-position of the resonances will be determined by the exact real atom-ion potential, a relative changes with the trap frequency will be roughly the same. The difference in the results for Li and He stem mainly from the different polarizability of the two species.   

\section*{Experimental considerations}

For experiments with single trapped ions, measuring the absolute ion-atom cross section is rather challenging, since there is only a single target particle and after each ion-atom collision event, the ion has to be re-initiated in its motional ground state. Similarly, determining ion mobilities would be a very challenging task. However, since the heating of a single ion out of its ground state can experimentally be measured with nearly 100$\%$ detection efficiency~\cite{Diedrich1989,Deslauries2006}, a better strategy to measure the consequence of the de-localized charge distribution of a ground state cooled ion in ion-atom collisions could be to measure the averaged energy transfer $\langle E_{\text{transfer}}\rangle$ to the ion. In cases where $\langle E_{\text{transfer}}\rangle$ is smaller than one motional quanta, its value should be rather easy to determine by measuring the rate at which an initially ground-state cooled ion is heated to an excited motional state using standard trapped-ion heating rate measurements based on so-called optical shelving ~\cite{Diedrich1989,Deslauries2006}. For much larger values of $\langle E_{\text{transfer}}\rangle$, thermometric methods based on measuring the strengths of various motional sidebands could be applied~\cite{Poulsen2012}, or alternatively, an extremely dilute atom gas may be applied, such that the heating rate would be dominated by distant but more frequent and less energy-changing collisions. For the ranges in Fig.~\ref{fig6}, where we expect the most pronounced effect on the de-location of the charge of an ion in the ground state, the latter approaches will probably have to be used since the averaged energy transfer in a single ion-atom collision here is much larger than a single quanta of motional energy. 

It should be noted that since the features in the average energy transfers are very broad with respect to the collisional energy, even with thermal ensembles of atoms, one should be able to clearly observe the change in the collisional properties due to differences in the de-localized charge distributions.

\section*{Conclusions}

Substituting a trapped ion by an extended charged distribution, according to its wavefunction, enables the treatment of trapped ion-atom interactions within a time-independent scattering framework. Using this approach, it is possible to study the role of the ionic trap on atom-ion scattering using a full quantum mechanical approach. The results show that the trapping frequency depends on a trapped ion's elastic and transport cross sections in an atomic neutral bath. Once the Langevin impact parameter matches the delocalization extension of the trapped ion, the elastic and transport cross sections show a variation of more than one order of magnitude within two decades of collision energy. Surprisingly enough, the observed deviations occur in the cold regime and beyond for collision energies $\gtrsim 1$mK. On the contrary, in the ultracold regime, the charge-induced dipole interaction dominates the trend of the scattering observables. However, fine details such as resonance positions and widths depend on the trapping frequency. 

Fueled by our results, we propose an experimental platform dedicated to measuring the explicit impact of the trapping potential on ion-atom scattering. The idea is to measure the energy transfer to the ion from the atoms, yielding a heating of the ion. The magnitude of the typical energy transfer after an ion-atom collision is generally larger or similar to the trap frequency, so this energy can be efficiently monitored in the experiment. Studying different atoms and trapping frequencies will explore the trap's role in the ion transport properties in the cold and ultracold regimes.

The results presented in this work on the role of the trap in the ion transport properties complements the  findings on trap-assisted ion-atom complexes formation and its impact on ionic polaron formation, where it has been shown that the trapping potential helps to accrete more atoms. Therefore, our findings help establish a new front and experimental platform to test novel features of the trapping potential. 

\section*{Methods}\label{method}

\subsection{Length scale of the charged distribution}

The charge distribution is given by

\begin{equation}
\rho(R)=e\left(\frac{1}{\sqrt{\pi} l_{\text{ion}}} \right)^{3}e^{-\frac{R^2}{l_{\text{ion}}^2}},
\end{equation}
that solely depends on the length scale associated to the ion in the harmonic oscillator potential $l_{\text{ion}}$. The average seize of the charge distribution $\langle R \rangle = l_{\text{ion}}/\sqrt{\pi}$ and its standard deviation is $\sigma_R=l_{\text{ion}}\sqrt{3/2-4/\pi}$. Based on these observations, we define 

\begin{equation}
l_{\text{distrib}}=\langle R \rangle +\sigma_R =l_{\text{ion}} \left( \frac{2}{\sqrt{\pi}} + \sqrt{\frac{3}{2}-\frac{4}{\pi}} \right), 
\end{equation}
as the characteristic length scale characterizing the charged distribution. It is worth noticing that this length scale determines the largest extension of the charge distribution with different behavior from the point charge case. Therefore, it is expected to be the dominant length scale in scattering problems at low collision energies.

\subsection{Scattering calculations}
The radial Schrödinger for the charged ion-atom system is given by
\begin{equation}
    \frac{1}{r}\frac{d^2}{dr^2}(r\psi_l) + (k^2-\frac{l(l+1)}{r^2}-\frac{2\mu V(r,\omega,m_{\text{ion}})}{\hbar^2})\psi_l=0, \label{eqS1}
\end{equation}
where $l$ is the angular momentum quantum number, $k=\sqrt{2\mu E_{\text{col}}}/\hbar$ is the wave vector, $\mu$ is the reduced mass of the atom-ion system, and $E_{\text{col}}$ is the collision energy. Eq.~\eqref{eqS1} is numerically solved using the Numerov algorithm~\cite{numerov} up to a final point of propagation $r_{\text{max}}$, where the solution is compared with its asymptotic solution 

\begin{equation}
    \psi_l(r\to \infty) = A_l(k)\frac{\sin(kr-l\pi/2+\delta_l)}{kr},
\end{equation}
where $\delta_l$ is the energy-dependent phase-shift. Finally, the elastic cross section is given by:
\begin{equation}
    \sigma_e(E_{\text{col}}) = \frac{4\pi}{k^2}\sum_{l=0}^{l_{max}} (2l+1)\sin^2(\delta_l).
\end{equation}


\subsection{Computational Details}
All the calculations were done in atomic units and with SciPy and NumPy packages~\cite{2020SciPy-NMeth,harris2020array}. Masses of Barium-137 ion, Lithium-6 atom and He-4 atom used were $250331.7953$, $10964.8925$, and $7296.2986 $ respectively. Polarizabilities of the Lithium-6 and Helium-4 atom were $164.1125$ and $1.4037$. The long range interaction coefficients used for Lithium-6 and Helium-4 were $100000$ and $4202.86$. The energy we sampled ranged from $1\mu$K to $100$K with 250 points, and for the wavefunction we scanned $500000$ points from $5.5$ to $12000$. For Helium atom, the maximum number of $l$ ranged form $10$ to $40$ and from $100$ to $4000$ for quantum and semi-classical calculations respectively. For Lithium atom, the maximum number of $l$ ranged form $40$ to $280$ and from $40$ to $5000$ for quantum and semi-classical calculations respectively.

\subsection{Semi-classical scattering}
Due to the strong nature of the charged-induced dipole interaction at almost any collision energy, many partial waves will contribute to the scattering, and hence it is possible to apply the semi-classical approximation to calculate the phase-shift, given by~\cite{landau2013quantum} (in atomic units)
\begin{equation}
    \delta_l^{SC} = -\int_{r_0}^{\infty}\frac{\mu V(r')}{\sqrt{k^2-\frac{l^2}{r'^2}}}dr',
\end{equation}
where $r_0=l/k$ is the classical turning point. Assuming a charged-induced dipole interaction $V(r)=-\frac{\alpha}{2r^4}$, and hence the semi-classical phase-shift reads as
\begin{equation}
    \delta_l^{SC}(E_{\text{col}}) = -\frac{\pi \mu^2 \alpha E_{\text{col}}}{4l^3},
\end{equation}
so the elastic cross section is given by
\begin{eqnarray}
\label{eq12}
\sigma_e^{SC}(E_{\text{col}}) &= & \frac{8\pi}{2\mu E_{\text{col}}}\int_{0}^{\infty}l \sin^2{\left(\delta_l^{SC}(E_{\text{col}})^2\right)}dl  \nonumber \\
 & = &  -\frac{\pi^{5/3}\alpha^{2/3}\mu^{1/3}\Gamma(-2/3)2^{1/3}}{6}E_{\text{col}}^{-1/3}. 
\end{eqnarray}
This is the results that we use for the point charged case along the main text.

\subsection{Energy transfer}

Let's assume that the atom is moving along the $z$ axis with a given momentum, $k$, and collision energy $E_{\text{col}}$. In this scenario, the energy transfer per collision energy is defined as the change in energy of the atom, and it is given by

\begin{equation}
\label{eq13}
\langle \Delta E_{\text{transfer}}\rangle =   \frac{\langle\mathbf{p}\cdot d\mathbf{p}\rangle}{\mu}=\frac{p^2}{\mu} \langle 1-\cos{(\theta)} \rangle,
\end{equation}
where $\langle x \rangle$ stands for the average of the observable $x$ along all possible scattering angles. Here, $\mathbf{p}$ represents the initial momentum of the colliding partners, $d\mathbf{p}$ the change of the momentum as consequence of the collision and 

\begin{equation}
\label{eqdiff}
    \langle 1-\cos{(\theta)} \rangle=\frac{\sigma_D(E_{\text{col}})}{\sigma_e(E_{\text{col}})}.
\end{equation}
In this equation the diffusion cross section (also known as the momentum transfer cross section) reads as
\begin{equation}
 \sigma_D(E_\text{col})=\int \frac{d\sigma_{el}(E_\text{col},\theta)}{d\Omega}(1-\cos{\theta})d\Omega
\end{equation}
where $\frac{d\sigma_{el}(E_\text{col},\theta)}{d\Omega}$ is the differential cross section for a scattering angle $\theta$ and $d\Omega=2\pi \sin{\theta}d\theta$, due to the azimuthal symmetry of the system under consideration. It is worth noticing that only when the differential cross is isotropic the diffusion and elastic cross section are the same. 

Using a classical capture model, it is possible to show that the diffusion cross section for ion-atom collision fulfills $\sigma_D(E_{\text{col}})\propto \frac{1}{\sqrt{E_{\text{col}}}}$, and bearing in mind that the elastic collision is $\propto E_{\text{col}}^{-1/3}$, as shown in Eq.~(\ref{eq12}), we find that 

\begin{equation}
\langle \Delta E_{\text{transfer}}\rangle \propto E^{5/6}.
\end{equation}

Another relevant transport cross section is the viscosity cross section, defined as

\begin{equation}
\label{eqvisco}
    \sigma_\eta(E_\text{col})=\int \frac{d\sigma_{el}(E_\text{col})}{d\Omega}(1-\cos^2{\theta})d\Omega,
\end{equation}
and it is essential to compute the viscosity of a fluid.

Any transport cross sections can be calculated quantum mechanically, yielding ~\cite{Mott}
\begin{equation}
\label{eq15}
    \sigma_D(E_\text{col})=\frac{4\pi}{k^2}\sum_{l=0}(l+1) \sin^2{(\delta_{l+1}(E_\text{col})-\delta_l(E_\text{col}))},
\end{equation}
for the diffusion cross section and
\begin{equation}
\label{eq16}
    \sigma_\eta(E_\text{col})=\frac{4\pi}{k^2}\sum_{l=0}\frac{(l+1)(l+2)}{2l+3} \sin^2{(\delta_{l+2}(E_\text{col})-\delta_l(E_\text{col}))},
\end{equation}
for the viscosity cross section. Eqs.~(\ref{eq15}) and (\ref{eq16}) have been used to calculate the viscosity and diffusion cross section presented in this work. The relevant phase-shifts were obtained from the elastic cross section calculation.

\section*{acknowledgements}
This work was supported by the United States Air Force Office of Scientific Research [grant number FA9550-23-1-0202]. J. P.-R. acknowledges the hospitality of the Department of Physics and Astronomy of Aarhus University, where some parts of this work was done.

\bibliographystyle{unsrt}
\bibliography{apssamp}

\end{document}